\documentclass[aps,prl,amsmath,amsfonts,superscriptaddress,twocolumn]{revtex4}
\usepackage{epsfig,graphicx}

\newcommand{\beq}{\begin{equation}}
\newcommand{\enq}{\end{equation}}

\begin{document}

\title{Dressed Feshbach molecules in the BEC-BCS crossover}

\author{M.W.J. Romans}

\affiliation{Institute for Theoretical Physics, Utrecht
University, Leuvenlaan 4, 3584 CE Utrecht, The Netherlands}

\author{H.T.C. Stoof}

\affiliation{Institute for Theoretical Physics, Utrecht
University, Leuvenlaan 4, 3584 CE Utrecht, The Netherlands}

\begin{abstract}
We present the RPA theory of the BEC-BCS crossover in an atomic
Fermi gas near a Feshbach resonance that includes the relevant
two-body atomic physics exactly. This allows us to determine the
probability $Z$ for the dressed molecules in the Bose-Einstein
condensate to be in the closed channel of the Feshbach resonance
and to compare with the recent experiments of Partridge {\it et
al.} [cond-mat/0505353] with $^{6}$Li. We determine for this
extremely broad resonance also the condensate density of the
dressed molecules throughout the BEC-BCS crossover.
\end{abstract}

\pacs{03.75.-b,67.40.-w,39.25.+k}

\maketitle

{\it Introduction.} --- The superfluid phase in an atomic Fermi
gas near a Feshbach resonance realizes a fundamentally new state
of matter, which shows a macroscopic coherence between atom pairs
and molecules that is controlled by the applied magnetic field. As
a result, such a gas offers the exciting possibility to study in
great detail the crossover between the Bose-Einstein condensation
(BEC) of diatomic molecules and the Bose-Einstein condensation of
atomic Cooper pairs, i.e., the Bardeen-Cooper-Schrieffer (BCS)
transition
\cite{stoof1996,timmermans2001,ohashi2002,milstein2002}. In fact,
the BEC-BCS crossover region is presently already actively being
explored by a number of experimental groups around the world
\cite{regal2004,MIT,Duke,Innsbruck,ENS,Rice}.

In more detail the physics of the BEC-BCS crossover occurring near
a Feshbach resonance can be understood as follows: The superfluid
phase of the gas is always associated with a Bose-Einstein
condensate of pairs, but the wave function of the pairs or dressed
molecules is given by the linear superposition
\begin{eqnarray}
\langle {\bf r}|\chi_{\rm dressed}\rangle = \sqrt{Z}\chi_{\rm
m}({\bf r}) |{\rm closed} \rangle + \sqrt{1-Z}\chi_{\rm aa}({\bf
r}) |{\rm open}\rangle~. \nonumber
\end{eqnarray}
In the BEC limit the applied magnetic field is taken such that the
bare molecular energy level lies far below the threshold of the
two-atom continuum and we have $Z \simeq 1$. In that case we are
dealing with a Bose-Einstein condensate of tightly-bound diatomic
molecules and the spatial part of the pair wave function is equal
to the bare molecular wave function $\chi_{\rm m}({\bf r})$. The
spin part of the pair wave function is then equal to $|{\rm
closed} \rangle$, i.e., the spin state of the closed channel that
causes the Feshbach resonance \cite{duine2003}. In the BCS limit
the bare molecular energy level lies far above the threshold of
the two-atom continuum and can be adiabatically eliminated. We
then have that $Z \simeq 0$ and the spatial part of the pair wave
function equals the usual BCS wave function for atomic Cooper
pairs $\chi_{\rm aa}({\bf r})$. This Cooper-pair wave function
depends also on the magnetic field, because the effective
attraction between the atoms after the adiabatic elimination of
the bare molecular state depends on the energy of that state. The
spin state of the Cooper pairs is, however, always equal to the
spin state $|{\rm open}\rangle$ of the open channel of the
Feshbach problem.

The probability $Z$ plays therefore a crucial role in the
description of the BEC-BCS crossover since it quantifies the
amount of coherence between the atom pairs and the bare molecules
in the gas. Unfortunately, however, the various theories
\cite{falco2004a,strinati,levin,mackie,ho,griffin,burnett,wetterich,stringari}
that are presently being used to understand the outcome of the
experiments are unable to accurately determine this quantity. This
comes about because, either $Z$ is assumed to be zero from the
outset, the many-body theory does not incorporate the two-body
Feshbach physics exactly, or the theory is able to determine only
the total number of bare molecules in the gas and thus requires an
additional assumption about the total number of dressed molecules
to extract $Z$. This situation is particularly unsettling because
of the recent $^{6}$Li experiments of Partridge {\it et al.}
\cite{Rice}, which have used the photodissociation rate to measure
the value of $Z$ throughout the crossover regime. In view of the
above situation it is pressing to develop an {\it ab initio}
many-body theory for the calculation of $Z$. How that may be
achieved is the main topic of this Letter.

{\it BEC-BCS crossover theory.} --- Introducing creation and
annihilation operators for the bare molecules and atoms, the
effective grand-canonical hamiltonian of the gas with chemical
potential $\mu$ becomes \cite{drummond,timmermans1999,duine2003}
\begin{align}
\label{hamiltonian} H&=\int d{\bf x}~ \psi^{\dagger}_{\rm m}({\bf
x})\left( -\frac{\hbar^2\mbox{\boldmath $\nabla$}^2}{4m}+
\delta-2\mu \right) \psi_{\rm m}({\bf x})
\\ &+ \sum_{\sigma=\uparrow,\downarrow} \int dx~
\psi^{\dagger}_{\rm \sigma}({\bf x}) \left(
-\frac{\hbar^2\mbox{\boldmath $\nabla$}^2}{2m}-\mu \right)
\psi_{\rm \sigma}({\bf x}) \nonumber
\\ &+g \int dx \left( \psi^{\dagger}_{\rm m}({\bf x})\psi_{\rm
\uparrow}({\bf x})\psi_{\rm \downarrow}({\bf
x})+\psi^{\dagger}_{\rm \downarrow}({\bf x})\psi^{\dagger}_{\rm
\uparrow}({\bf x})\psi_{\rm m}({\bf x}) \right) \nonumber
\\ &+ \frac{4\pi a_{\rm bg} \hbar^2}{m} \int dx~ \psi^{\dagger}_{\rm
\downarrow}({\bf x})\psi^{\dagger}_{\rm \uparrow}({\bf
x})\psi_{\rm \uparrow}({\bf x})\psi_{\rm \downarrow}({\bf x})~,
\nonumber
\end{align}
where the two hyperfine states of the atoms are denoted by $|
\uparrow\rangle$ and $|\downarrow\rangle$, and the magnetic-moment
difference $\Delta\mu_{\rm mag}$ between the hyperfine states
$|{\rm closed}\rangle$ and $|{\rm open}\rangle \equiv
(|\uparrow\downarrow\rangle-|\downarrow\uparrow\rangle)/\sqrt{2}$
gives the so-called detuning from resonance $\delta =
\Delta\mu_{\rm mag}(B-B_0)$. Note that the atom-molecule coupling
constant $g$ and the background scattering length $a_{\rm bg}$
depend on the magnetic field $B$ in such a manner that the total
scattering length $a = a_{\rm bg} - m g^2/4\pi\hbar^2\delta$
agrees with the Feshbach resonance of interest. This is especially
important for the broad Feshbach resonance near $834$ Gauss that
is used in all $^{6}$Li experiments up to date
\cite{houbiers,thomas}.

From now on we consider only the most interesting region close to
resonance, where $a_{\rm bg} \ll a$ and the effective interaction
between the atoms is dominated by the resonant part $-g^2/\delta$.
This suggests that the last term in the right-hand side of
Eq.~(\ref{hamiltonian}) can be neglected altogether. This is,
however, not true because we can neglect this term only after we
have included its effect on the atom-molecule coupling
\cite{falco2005}. Physically, the reason for this subtlety is that
the above Hamiltonian is an effective Hamiltonian that is only
valid for low energies. However, to properly account for the
two-body physics near a Feshbach resonance also high-energy states
are needed. The main effect of these high-energy states is to
renormalize the atom-molecule coupling to $g({\bf k}) \equiv
g/(1+ika_{\rm bg})$, where ${\bf k}$ is the relative momentum of
the atoms involved in the coupling \cite{duine2003,falco2005}.
Only after having performed this substitution are we allowed to
neglect the background interaction between the atoms.

Without the background interaction the atomic part of the
Hamiltonian is quadratic. Using standard functional methods the
fermionic fields can thus be integrated out exactly. This leads to
an effective action for the molecules that at sufficiently low
temperatures has a minimum at a nonzero value of $\langle
\psi_{\rm m}({\bf x}) \rangle \equiv \sqrt{Zn_{\rm mc}}$, where we
introduced the dressed molecular condensate density $n_{\rm mc}$.
Neglecting fluctuations at this point leads to a mean-field theory
of the BEC-BCS crossover. As mentioned in the introduction,
however, this mean-field theory is unable to calculate the
probability $Z$ since it only determines the bare molecular
condensate density $|\langle \psi_{\rm m}({\bf x}) \rangle|^2$. We
therefore also consider the quadratic fluctuations around the
minimum of the effective molecular action, i.e., we consider the
Bogoliubov theory of the bare molecules.

As expected with a spontaneously broken $U(1)$ symmetry associated
with the presence of a Bose-Einstein condensate, the gaussian
fluctuations are determined by normal and anomalous self energies
of the bare molecules, which at zero temperature reduce to
\begin{widetext}
\begin{eqnarray}
\hbar\Sigma_{11}({\bf k},i\omega) &=&
     \int \frac{d{\bf k'}}{(2\pi)^3}~|g({\bf k'})|^2
     \Bigg\{\frac{|u_a({\bf k}'_+)|^2 |u_a({\bf k}'_-)|^2}
       {i\hbar\omega-\hbar\omega_a({\bf k}'_+)-\hbar\omega_a({\bf k}'_-)}-
     \frac{|v_a({\bf k}'_+)|^2 |v_a({\bf k}'_-)|^2}
       {i\hbar\omega+\hbar\omega_a({\bf k'}_+)+\hbar\omega_a({\bf k}'_-)}+
     \frac{1}{2 \epsilon({\bf k}')}\Bigg\}~, \nonumber\\
\hbar\Sigma_{12}({\bf k},i\omega) &=& 2
     \int \frac{d{\bf k}'}{(2\pi)^3}~|g({\bf k}')|^2 \Bigg\{
       u_a({\bf k}'_+)v_a({\bf k}'_+)u_a({\bf k}'_-)v_a({\bf k}'_-)
     \frac{\hbar\omega_a({\bf k}'_+)+\hbar\omega_a({\bf k}'_-)}
       {(\hbar\omega_a({\bf k}'_+)+\hbar\omega_a({\bf k'}_-))^2
           +(\hbar\omega)^2}\Bigg\}~.
\end{eqnarray}
\end{widetext}
Here we have also introduced the BCS dispersion $\hbar\omega_{\rm
a}({\bf k})=\sqrt{(\epsilon({\bf k})- \mu)^2 + |g({\bf
k})|^2Zn_{\rm mc}}$, the bare atomic dispersion $\epsilon({\bf
k})=\hbar^2{\bf k}^2/2m$, the usual BCS coherence factors $u_{\rm
a}({\bf k})$ and $v_{\rm a}({\bf k})$, and the notation ${\bf
k}'_{\pm}={\bf k}/2 \pm {\bf k}'$. In terms of the above self
energies the minimum of the effective action is determined by the
exact Hugenholtz-Pines relation $2\mu = \delta +
\hbar\Sigma_{11}({\bf 0},0)- \hbar\Sigma_{12}({\bf 0},0)$, which
turns out to be equal to a modified BCS gap equation
\begin{eqnarray}
\delta - 2\mu = \int \frac{d{\bf k}}{(2\pi)^3}~ |g({\bf k})|^2
\left( \frac{1}{2\hbar\omega({\bf k})} - \frac{1}{2\epsilon({\bf
k})} \right).
\end{eqnarray}
Finally, we also need the equation of state, which we for
consistency reasons \cite{duine2003} obtain by differentiating the
thermodynamic potential with respect to the chemical potential.
Including the effect of the fluctuations we obtain for the total
density of atoms
\begin{equation}
\label{eqstate} n = - {\rm Tr}[G_{\rm a}] + 2Zn_{\rm mc} - {\rm
Tr}[G_{\rm m}] + \frac{1}{2} {\rm Tr}\left[G_{\rm m}
\frac{\partial\hbar\Sigma}{\partial\mu} \right]~,
\end{equation}
where $G_{\rm a}$ and $G_{\rm m}$ are the Nambu ($2 \times
2$)-matrix Green's functions of the bare atoms and molecules,
respectively. For a given density and magnetic field the last two
equations determine only the bare molecular condensate density and
the chemical potential. Hence, we need a third equation to
determine also $Z$.

{\it Dressed molecules.} --- Before we derive this missing
equation, let us first discuss in some more detail the physics
behind the maybe somewhat unexpected equation of state in
Eq.~(\ref{eqstate}). The first two terms represent the mean-field
theory without fluctuations that is most often used in the recent
literature \cite{levin,mackie,ho,burnett}. Because of the absence
of fluctuations all the molecules are Bose-Einstein condensed and
there is no depletion. The third term precisely corresponds to
this depletion. Finally, the fourth term physically describes the
dressing of the bare atoms and molecules. This can be made more
clear by reformulation the equation of state in terms of dressed
atoms and molecules, instead of bare atoms and molecules. Since
every dressed molecule contains two atoms, we expect the
contribution $2n_{\rm mc}$ from the condensate of dressed
molecules. Indeed, in the BEC limit it can be shown explicitly
that the atomic density $- {\rm Tr}[G_{\rm a}] = 2 \int d{\bf k}
|v_{\rm a}({\bf k})|^2/(2\pi)^3$ contains exactly the expected
contribution $2(1-Z)n_{\rm mc}$ of paired atoms in the
Bose-Einstein condensate of dressed molecules. The atomic density
does, however, not contain the paired atoms associated with the
dressed molecules that are not in the Bose-Einstein condensate.
This omission is repaired by the fluctuation corrections which
contain both the associated changes in the atomic density and
twice the total density of dressed molecules that are not
Bose-Einstein condensed, i.e., twice the dressed molecular
depletion.

\begin{figure}[t]
\epsfig{figure=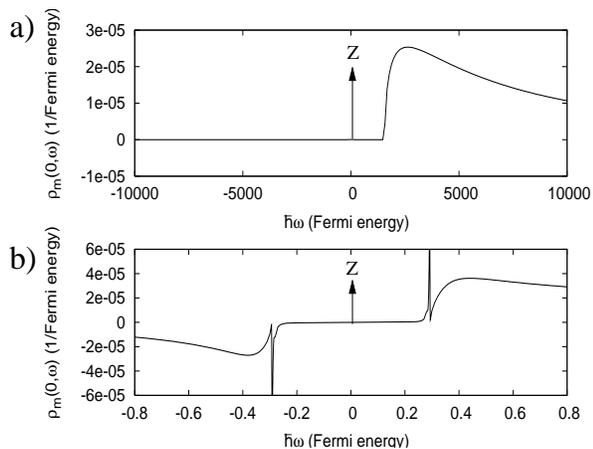,width=7.8cm} \caption{\rm Spectral
functions of the bare $^6$Li$_2$ molecules with zero momentum a)
in the BEC limit at $571$ Gauss and b) in the BCS limit at $892$
Gauss of the crossover occurring near the broad Feshbach resonance
of $^6$Li at $834$ Gauss. The Fermi energy of the gas is $380$ nK.
\label{spectralfunction} }
\end{figure}

We are now in a position to determine $Z$. In principle it is
equal to the residue of the pole of $G_{{\rm m};11}({\bf
0},i\omega)$ at $\omega=0$. To understand the physics of that
result better it is useful to consider the spectral function of
the bare molecules with zero momentum, i.e., $\rho_{\rm m}({\bf
0},\omega)= - {\rm Im}[G_{{\rm m};11}({\bf 0},\omega+i0)]/\pi$.
This spectral function is closely related to the density of states
of the molecules at zero momentum and thus gives detailed
information on the dressed molecular content of the gas. Most
importantly for our purposed, the Bose-Einstein condensate of
dressed molecules gives rise to a delta-function in the spectral
function exactly at zero frequency. The strength of this delta
function is precisely $Z$, because this is the probability to take
a bare molecule out of the Bose-Einstein condensate of dressed
molecules. Besides this bound-state contribution, the spectral
function contains also a contribution from the continuum of atomic
scattering states. In the BEC limit of the crossover the continuum
contribution only occurs at positive frequency and starts at a
frequency of $-2\mu/\hbar \simeq 2\hbar/ma^2$ as shown in Fig.~1a.
In the BCS limit the continuum contribution occurs both at
positive and negative frequencies, which start at a frequency of
about $\pm 2g\sqrt{Zn_{\rm mc}}/\hbar$, respectively, due to the
gap that exists for the creation of an atomic
quasiparticle-quasihole pair. This is shown in Fig.~1b. The
negative frequency part of the spectral function is especially
important, because it determines the depletion of the
Bose-Einstein condensate of dressed molecules. Physically it
represents the dressed molecules that are stabilized by the Fermi
sea \cite{falco2004b}. Making use of the above physical picture,
we finally obtain the desired result,
\begin{eqnarray}
Z = \frac{1-\Sigma^{(1)}_{11}}
    {\left(1-\Sigma^{(1)}_{11}\right)^2
    +\Sigma_{12}\left(\Sigma^{(2)}_{12}-\Sigma^{(2)}_{11}\right)}~,
\end{eqnarray}
where $\Sigma^{(n)}_{ij}\equiv(-i)^n\partial^n\Sigma_{ij}({\bf
0},0)/\partial\omega^n$.

At nonzero momenta the spectral function is similar but now
contains two delta functions at the frequencies $\pm \omega_{\rm
m}({\bf k})$, which have the strength $Z|u_{\rm m}({\bf k})|^2$
and $-Z|v_{\rm m}({\bf k})|^2$, respectively, with $|u_{\rm
m}({\bf k})|^2-|v_{\rm m}({\bf k})|^2=1$. This shows explicitly
how at long wavelengths our theory leads to a Bogoliubov-like
theory for dressed molecules with a wave function renormalization
factor $Z$. Moreover, in agreement with the Goldstone theorem, the
quasiparticle dispersion $\omega_{\rm m}({\bf k})$ always turns
out to be linear at long wavelengths. In the following we
therefore determine also the associated speed of (second) sound
throughout the BEC-BCS crossover region.

\begin{figure}[h]
\epsfig{figure=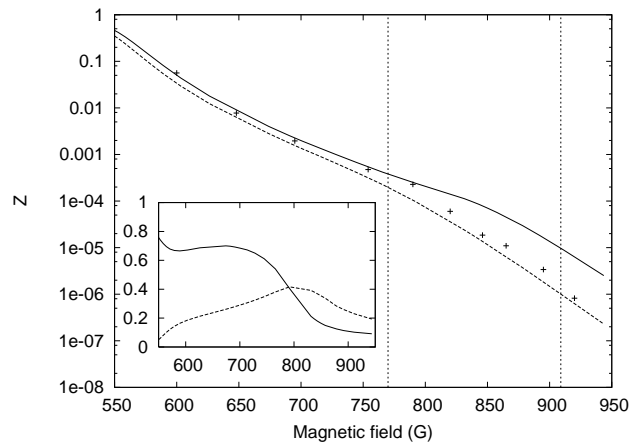,width=8.5cm} \caption{\rm The solid
curve shows the probability $Z$ and the dashed curve the fraction
$2Zn_{\rm mc}/n$ as a function of magnetic field. The Fermi energy
$\hbar^2k_{\rm F}/2m$ of the gas is $380$ nK. The data points are
from the experiment of Partridge {\it et al.} \cite{Rice}. In the
inset the solid line shows the Bose-Einstein condensate fraction
of dressed molecules $2n_{\rm mc}/n$ and the dashed line the
contribution of the fluctuations to the total atomic density. The
vertical lines indicate the magnetic fields where $k_{\rm
F}|a|=1$. \label{Z} \vspace*{-0.5cm}}
\end{figure}

{\it Results and discussion.} --- In Fig.~2 we show our results
for $Z$ and compare with the experimental data of Partridge {\it
et al.} \cite{Rice}. In general the agreement is satisfactory.
This is particularly true at relatively low magnetic fields where
$Z$ is determined by two-body physics, which is exactly
incorporated into our theory. At higher magnetic fields the
theoretical values of $Z$ are somewhat higher than the
experimental ones. We believe that the reasons for this are
twofold. First, the experiment is performed in an optical trap. As
a result the experimental data involves an appropriate average
over the density profile of the gas, which lowers the observed
value of $Z$. Second, the photodissociation laser used in the
experiment has a width which is much larger than the Fermi energy
of the gas. The laser, therefore, has not sufficient resolution to
probe only the Bose-Einstein condensate of dressed molecules, and
probes also the dressed molecules which are not Bose-Einstein
condensed. This second effect should be especially important at
high magnetic fields. To disentangle these different effects,
however, goes beyond the scope of the present Letter and is left for further investigation.

We also show in Fig.~2 the Bose-Einstein condensate fraction of
dressed molecules $2n_{\rm mc}/n$ throughout the BEC-BCS
crossover. In qualitative agreement with the poor man's
approach of Ref.~\cite{falco2004a}, the latter fraction is always
substantial below the Feshbach resonance and becomes negligible
only sufficiently far above the Feshbach resonance when $k_{\rm
F}|a|<1$. This is an important observation, because in our
theoretical description of the experiment of Partridge {\it et
al.} the molecular probe couples directly to the dressed
molecules, which act as distinct entities in the gas since the
atom-molecule coupling is much larger than the coupling of the
probe laser to the bare molecules. In this manner it is most easy
to understand the experimental observation that there is initially
an exponential (one-body) decay of a large part of the total
atomic density with a rate that is much smaller than the bare
molecular photodissociation rate.

\begin{figure}[h]
\epsfig{figure=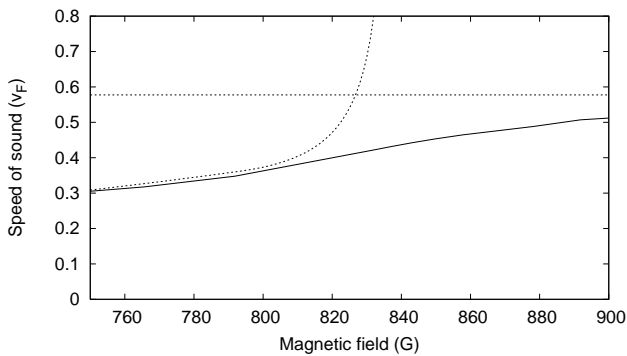,width=8.5cm} \caption{\rm The speed of
sound as a function of magnetic field. The Fermi energy of the gas
is $380$ nK. The dashed curves show the Bogoliubov result $\hbar
\sqrt{2\pi a n_{\rm mc}}/m$ and the speed of sound $\hbar k_{\rm
F}/ \sqrt{3}m \equiv v_F/\sqrt{3}$ of the Anderson-Bogoliubov mode.
\label{scatteringlength}}
\end{figure}

For completeness we give in Fig.~2 also the Bose-Einstein
condensate fraction of bare molecules $2Zn_{\rm mc}/n$ and the
fluctuation corrections to the total atomic density. As expected
the fluctuation corrections are very important in the crossover
region and become small far away from the Feshbach resonance,
where mean-field theory applies. From the fluctuations we also
extract the speed of sound of the gas, which is shown in
Fig.~3. Note that in the BCS limit the Anderson-Bogoliubov mode is
recovered. In combination with the presence of the sharp peaks in the spectral function in Fig.~1b,
this shows that also the decoupling of the amplitude and phase
fluctuations of the Bose-Einstein condensate of dressed molecules
that occurs in the BCS limit is correctly incorporated. We
therefore conclude that the RPA-like atom-molecule theory
developed here gives an excellent account of the subtle interplay
between two-body and many-body physics taking place at the
crossover near a Feshbach resonance.

We are very grateful for many helpful remarks and stimulating
discussions with Randy Hulet. This work is supported by the
Stichting voor Fundamenteel Onderzoek der Materie (FOM) and the
Nederlandse Organisatie voor Wetenschaplijk Onderzoek (NWO).

\bibliographystyle{apsrev}

\end{document}